\documentclass[twocolumn,prl,nocentre]{revtex4}% Physical Review B
\usepackage{graphicx}% Include figure files
\usepackage{dcolumn}% Align table columns on decimal point
\usepackage{bm}% bold math
\usepackage{amsmath} %[tbtags] or [centertags] options to tune where eq. tag is placed (last line or centered)
\usepackage{subfigure}
\usepackage{MnSymbol} % for extra math symbols like \llangle (<<)

\usepackage{xcolor}

\begin{document}
\title{Single-layer Skyrmions in a van der Waals antiferromagnet}
\author{A. Pozzi}
\thanks
{Corresponding author} 
\email{a.pozzi@rug.nl}
\author{M. Mostovoy}
\affiliation{Zernike Institute for Advanced Materials, 
	University of Groningen, Nijenborgh 4,9747 AG Groningen,  The Netherlands}

\date{\today}

\begin{abstract}
{
Transition metal halides are quasi-two-dimensional van der Waals magnets that can potentially host antiferromagnetic (AFM) skyrmions induced by competing exchange interactions.
We study theoretically magnetic states of Fe-doped NiBr$_2$ using spin interactions obtained by fitting experimental data.
We find that AFM interlayer interactions suppress skyrmion lattices but allow for skyrmions with topological charges $1,2$ and $3$ confined to a single magnetic layer.
These single-layer skyrmions exist in three different magnetic phases and their dynamics changes drastically
at the transition from a uniform to a non-collinear modulated state.  
}
\end{abstract}

% %%% PACS numbers

%\keywords{ }%Use showkeys class option if keyword
% display desired
         
\maketitle

% \clearpage
%

%\vspace{5mm}

%%INTRODUCTION
%{\flushleft{\em Introduction}: }
Since their first observation \cite{Muelbauer2009,Yu2010}, magnetic skyrmions are in the focus of many theoretical and experimental studies. 
Electrically driven dynamics and stability rooted in non-trivial topology make skyrmions promising information carriers in new magnetic memory devices \cite{Nagaosa2013,Fert2013,Fert2017,Back2020}. 
Current research on skyrmions is focused on magnets with chiral crystal lattices and multilayer devices where inversion symmetry is broken at interfaces between magnetic and heavy-metal materials \cite{Jiang2015,Woo2016}.  
The lack of inversion leads to anisotropic spin interactions favoring non-collinear magnetic textures \cite{Bogdanov1989}.

Inversion symmetry breaking is not strictly necessary: the competing Heisenberg exchange interactions in Mott insulators  \cite{Okubo2012,Leonov2015,Batista2016} and long-ranged spin interactions mediated by itinerant electrons in magnetic conductors \cite{Azhar2017,Hayami2017, Ozawa2017,Okumura2020} can stabilize skyrmions and even more complex spin textures, such as hedgehog crystals \cite{Ishiwata2018}.
Tiny (2-3 nm in diameter) skyrmions recently found in  centrosymmetric intermetallics give rise to large Topological Hall and Nernst effects \cite{Kurumaji2019,Hirshberger2019,Khanh2020,Hirshberger2020}.  
Isotropic interactions in achiral magnets result in additional low-energy skyrmion degrees of freedom, helicity and voriticity, which strongly modify interactions and dynamics of these topological defects \cite{Leonov2015,Leonov2017, Kharkov2017}.  
There is also a lot of interest in skyrmions in natural and artificial AFM materials that may propagate with a higher speed  and show no Skyrmion Hall effect  \cite{Barker2016,Zhang2016,Legrand2020}. 

Here, we explore the possibility to find skyrmions in layered transition metal halides, which are wide gap Mott insulators showing a variety of magnetic orders \cite{McGuire2017}. 
These van der Waals materials allow for exfoliation down to monolayers, which enables control of two-dimensional magnetism by doping, gating, magnetic and electric fields \cite{Jiang2018}. 
We focus on NiBr$_2$ that undergoes a transition into the collinear  A-type AFM state at $T_{\rm N} = 52$ K, which transforms into a multiferroic spiral state below 23 K with six energetically equivalent orientations of the in-plane wave vector \cite{Day1976,Adam1980,Tokunaga2011}.
The spin spiral results from the competition between the nearest-neighbor ferromagnetic (FM) Heisenberg exchange interactions on the $\sim 90^\circ$ metal-halogen-metal bonds and  further-neighbor AFM interactions enhanced by strong  metal-ligand covalency and lattice geometry.
The sign of magnetic anisotropy can be controlled  in  this material by doping \cite{Moore1985}.

Our study of a model of doped NiBr$_2$ with parameters obtained by fitting experimental data shows that the AFM interlayer interactions suppress skyrmion crystals at all magnetic fields and concentrations of dopands, but allow for topological defects with skyrmion topology, which are trully two-dimensional as their topological charge is only nonzero in a single magnetic layer.
These single-layer skyrmions and bi-merons are stable in three different magnetic phases at experimentally accessible magnetic fields.
We discuss dynamics of these defects induced by the spin-Hall torque. 

%\section*{Results}
{\flushleft\it The model.}
The spin Hamiltonian reads, 
\begin{equation}
\label{eq:model}
\begin{aligned}
H =
&-J_1\sum_{\left<i,j\right>,\lambda} \mathbf{S}^{\lambda}_i\cdot\mathbf{S}^{\lambda}_j
+J_2\sum_{\left\llangle i,j \right\rrangle,\lambda} \mathbf{S}^{\lambda}_i\cdot\mathbf{S}^{\lambda}_j\\
&+\sum_{i,\lambda} \left[J_{\perp}\mathbf{S}^{\lambda}_{i}\cdot\mathbf{S}^{\lambda+1}_i 
-\frac{K}{2}\left(\mathrm{S}^{\lambda}_{i}\right)_{z}^{2}
- \mathbf{h} \cdot \mathbf{S}_{i}^{\lambda}\right] ,
\end{aligned}
\end{equation}
where $\mathbf{S}^\lambda_i$ denotes the unit spin vector at the site $i$ of a triangular lattice in the layer $\lambda$. 
Indices in single and double brackets denote pairs of nearest-neighbor (NN) and next-nearest-neighbor (NNN) sites in the triangular lattice, respectively. 
The NN Heisenberg exchange interaction is ferromagnetic, whereas the NNN exchange that gives rise to frustration and the interlayer interaction are antiferromagnetic ($J_1,J_2,J_\perp > 0$); $K$ is the single-ion anisotropy and $\mathbf{h}$ is the magnetic field. 
%
%All model parameters are measured in units of $J_1 = 1$.  

Equation (\ref{eq:model}) contains a few simplifications. 
First, NiBr$_2$ has a CdCl$_2$ crystal structure (space group $R\bar{3}m$)  with ABC stacking of three triangular layers in the rhombohedral unit cell \cite{McGuire2017}.
Second, the third-nearest-neighbor AFM exchange  in NiBr$_2$ is stronger than the NNN one \cite{Regnault1982,Day1984}.
Our model is justified by the long period of the spiral modulation ($\sim 18$ in-plane  lattice constants), which makes possible to ignore the relative shifts of neighboring layers.
Spin models with 2 and 3 exchange interactions become equivalent in the continuum limit, the only difference being the orientation of the spiral wave vector. 
\begin{figure}[htbp]
\centering{}\includegraphics[width=\columnwidth]{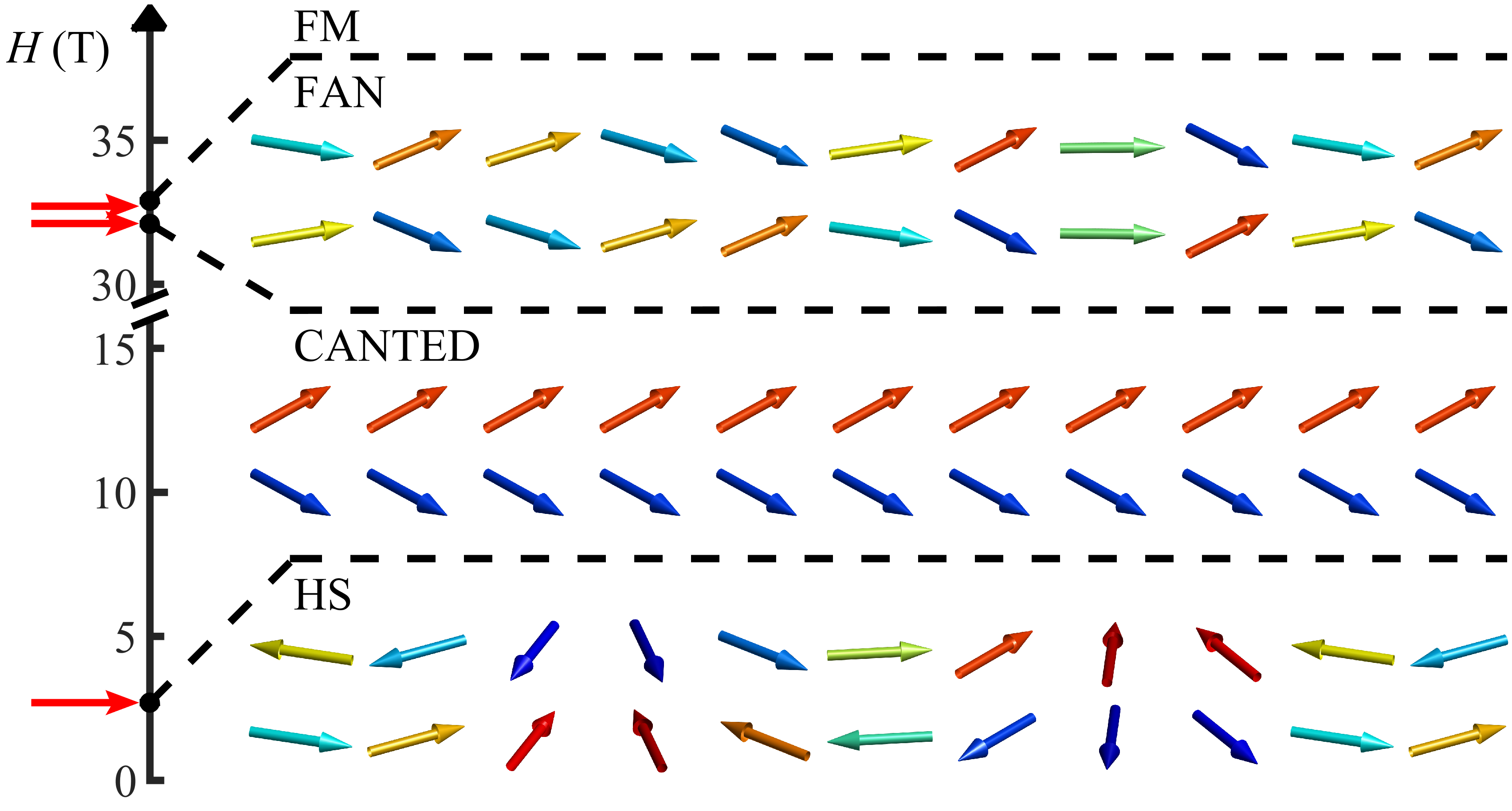}
\begin{centering}
\caption{
\label{fig:inplanephd}
Zero-temperature magnetic phase diagram of NiBr$_2$ in the magnetic field $H \perp c$. The black dots show the experimentally measured values of the
three critical fields, $H_1 < H_2 < H_3$, corresponding to the transitions between the spiral, canted AFM, fan, and ferromagnetic spin states
\cite{Katsumata1983}. Red arrows indicate the critical fields obtained in our numerical simulations, for $J_1 = 59.5 \,\mathrm{K}$, $J_2 = 20.4 \,\mathrm{K}$ and $J_\perp = 11.0 \,\mathrm{K}$. Spin configurations in two neighboring $ab$ layers are schematically shown with  arrows. 
}
\par\end{centering}
	
\end{figure}
\begin{figure}[htbp]
\centering{}\includegraphics[width=\columnwidth]{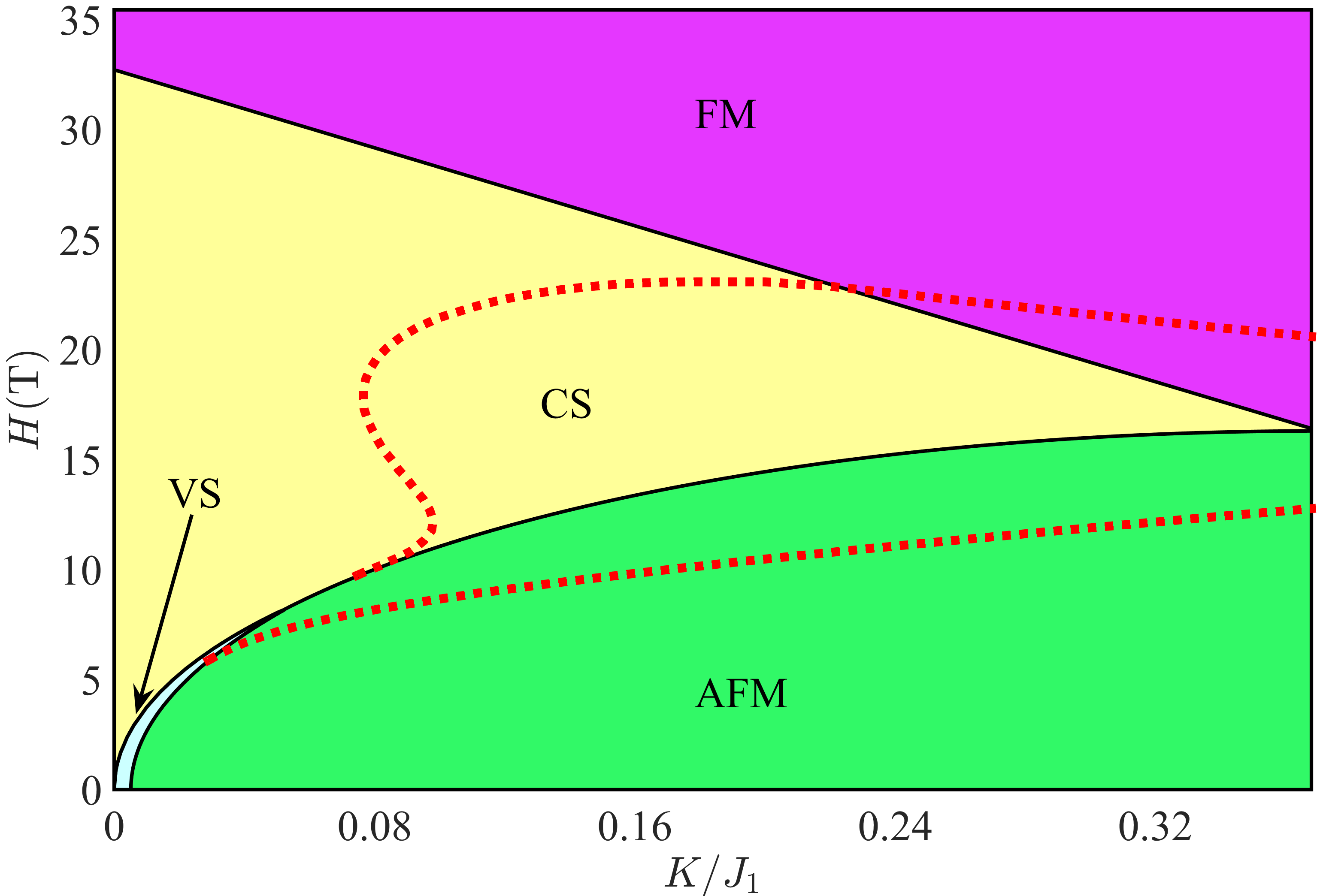}
	\begin{centering}
		\caption{
		\label{fig:PHD}
Magnetic field vs anisotropy phase diagram of 
Fe$\bm{_{x}}$N$\bm{_{1-x}}$Br$\bm{_2}$ for $H \| c$, which
includes the A-type AFM,  
vertical spiral (VS), conical spiral (CS) and  field-induced ferromagnetic (FM) states. The area encircled by red dotted line is the region 
of stability of defects with skyrmion charge $\pm 1$ confined to a single spin layer. 
}
 		
		\par\end{centering}	
\end{figure}
Earlier theoretical study of a single-layer frustrated triangular magnet with an easy-axis anisotropy ($K > 0$)  showed that the skyrmion crystal occupies a large part of the phase diagram  \cite{Leonov2015}. 
NiBr$_2$ is an easy-plane antiferromagnet ($K < 0$) \cite{Regnault1982,Day1984}, but its anisotropy changes sign upon 
substitution of magnetic ions:   
$\mathrm{Fe_{x}Ni_{1-x}Br_{2}}$ becomes an easy-axis magnet for $x>9.9\%$ \cite{Moore1985}.
AFM interlayer interactions in quasi-two-dimensional materials govern symmetry of skyrmion crystals and can suppress them \cite{LinBatista2018}.   
The question arises whether the interlayer interactions in dihalides allow for skyrmions. 
To address this question, we determine exchange parameters of our model using experimental data on NiBr$_2$.
With $J_2/J_1 = 0.343$ we fit the magnitude of the in-plane component of the spiral wave vector,   $\mathbf{Q} = [0.027,0.027,3/2]$ \cite{Moore1985}
(see Supplementary Information). 
The interlayer coupling $J_\perp$ is obtained from the fit of the phase diagram of NiBr$_2$ under an in-plane magnetic field \cite{Katsumata1983}, which shows three phase transitions: from the spiral to the canted AFM state at $H_1=2.7\,\mathrm{T}$, from the AFM to the fan state at $H_2=31.5\,\mathrm{T}$ and from the fan to saturated state at $H_3=32.9\,\mathrm{T}$ measured at 1.3 K (see Fig.~\ref{fig:inplanephd}).  
The $\frac{H_1}{H_3}$ ratio is a measure of the ratio of the exchange energy difference between the uniform AFM and spiral states in zero field, $\Delta E_{\rm ex}$, and the interlayer exchange energy:
\begin{equation}
\label{eq:EexJperp}
\frac{\Delta E_{\rm ex}}{J_\perp}\simeq \left(\frac{H_1}{H_3}\right)^2 \approx 6.7 \cdot 10^{-3}.
\end{equation}
Figure~\ref{fig:inplanephd} shows the comparison between the critical fields obtained in our simulations for $J_\perp / J_1 = 0.184$ (red arrows) and experimentally measured values (black dots). 
We converted $h$ into $H$ measured in Tesla using $H = \frac{J_1}{2 \mu_{\rm B}} h$ with $J_1 = 59.5 \,\mathrm{K}$, $ 2 \mu_{\rm B}$ being the magnetic moment of Ni$^{2+}$ ion. 
Since spins in all three magnetic phases are confined to the $ab$ plane, the critical fields are independent of $K$.

{\flushleft\it Magnetic states and topological defects.}
Next we calculate the $(H,K)$ phase diagram of the Fe-doped compound, Fe$_{x}$N$_{1-x}$Br$_2$, treating the anisotropy parameter $K$ as a variable that depends on the concentration of Fe dopands $x$. 
Figure~\ref{fig:PHD} shows the phase  diagram for $K > 0$ (easy-axis anisotropy) and $H \| c$ including the vertical spiral (VS), A-type AFM, conical spiral (CS) and the field-polarized (FM) phases. 
The small energy difference between non-collinear and collinear spin orders  [see Eq.(\ref{eq:EexJperp})] confines the VS state  to a narrow region  between the A-type AFM state with alternating layers of spins parallel/antiparallel to the $c$ axis and the CS phase, in which $S^z$ is the same in all layers and the sign of the in-plane spiral component alternates from layer to layer.  
The skyrmion lattices, both the triangular array of skyrmion tubes and a three-dimensional HCP/FCC crystal of skyrmions \cite{LinBatista2018}, are suppressed for the realistic model parameters alongside with all multiply-periodic magnetic phases found a single triangular layer \cite{Okubo2012,Leonov2015}.

\begin{figure}[htbp]

	\raggedright{\bf\large a \hskip 4.cm b}\\
	\includegraphics[height=0.49\columnwidth]{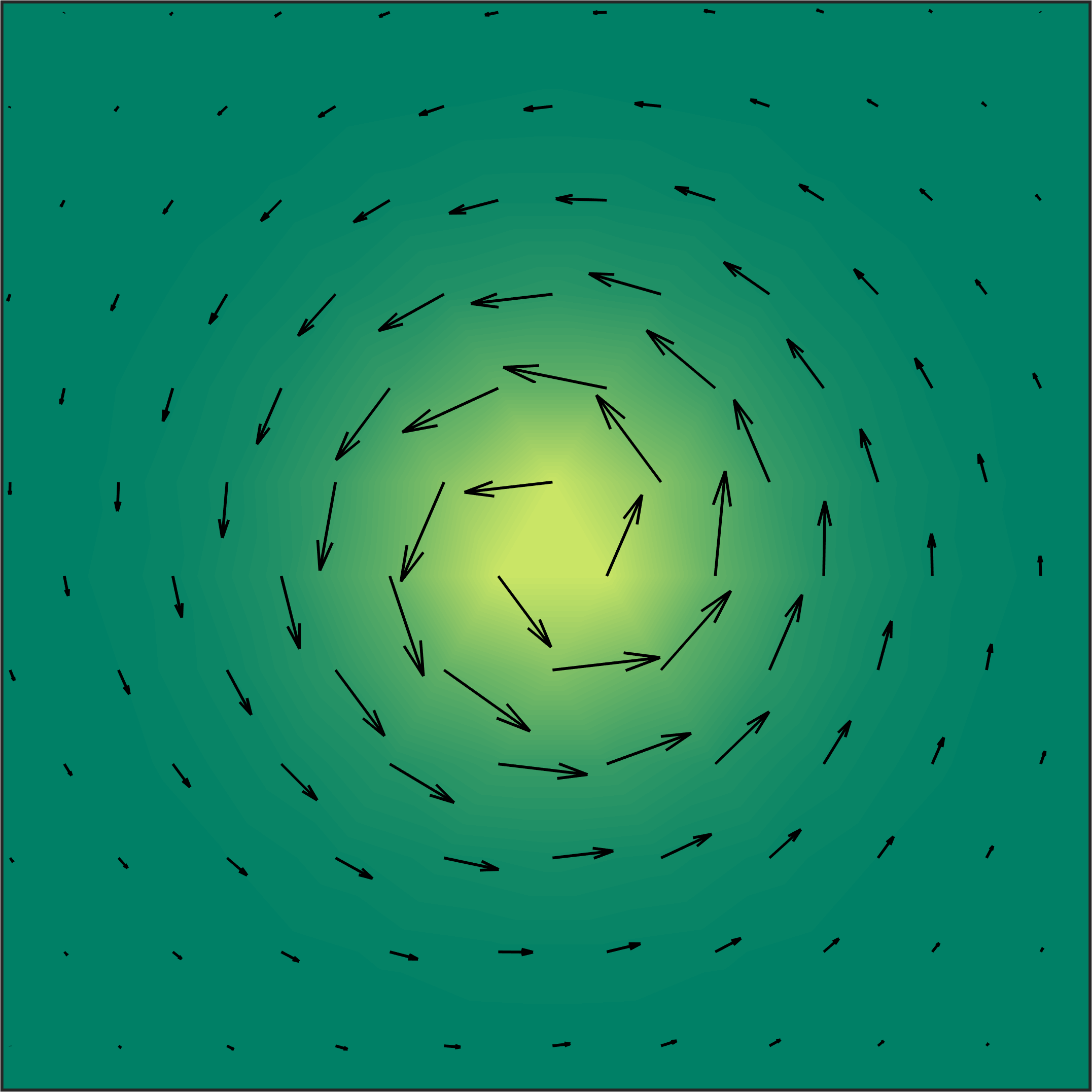}\hfill
	\includegraphics[height=0.49\columnwidth]{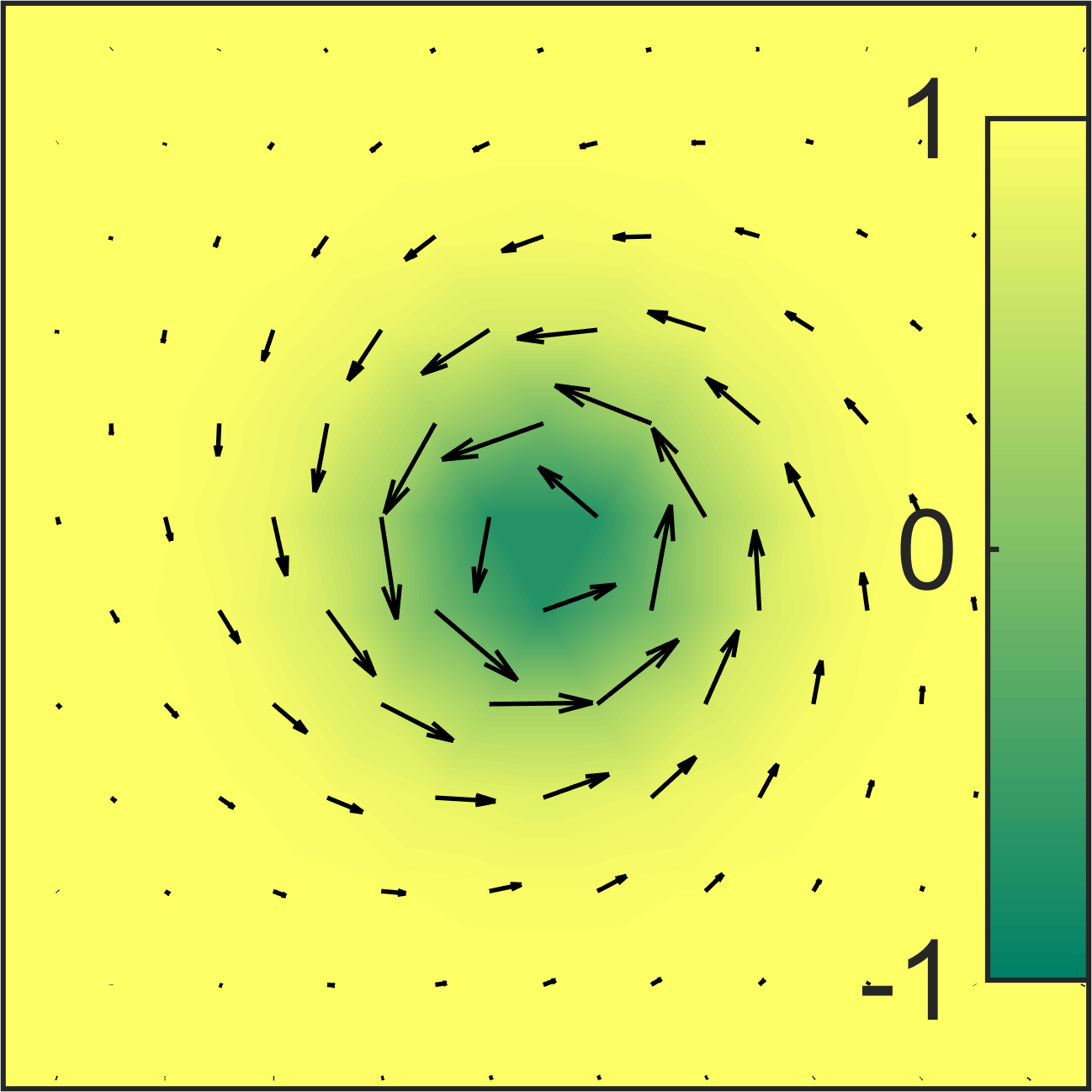}\\
	\raggedright{\bf\large c \hskip 4.cm d}\\
	\includegraphics[height=0.49\columnwidth]{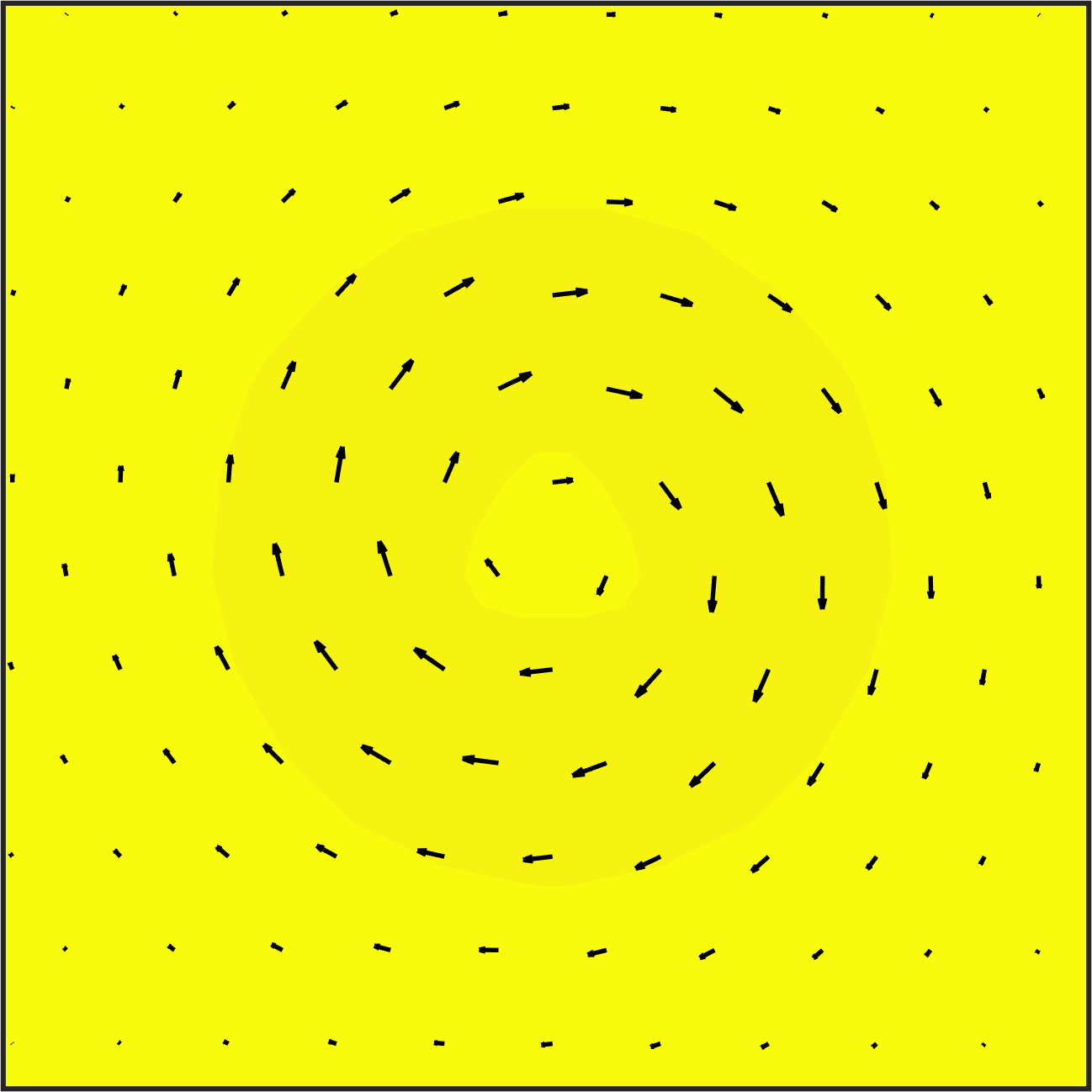}\hfill
	\includegraphics[height=0.49\columnwidth]{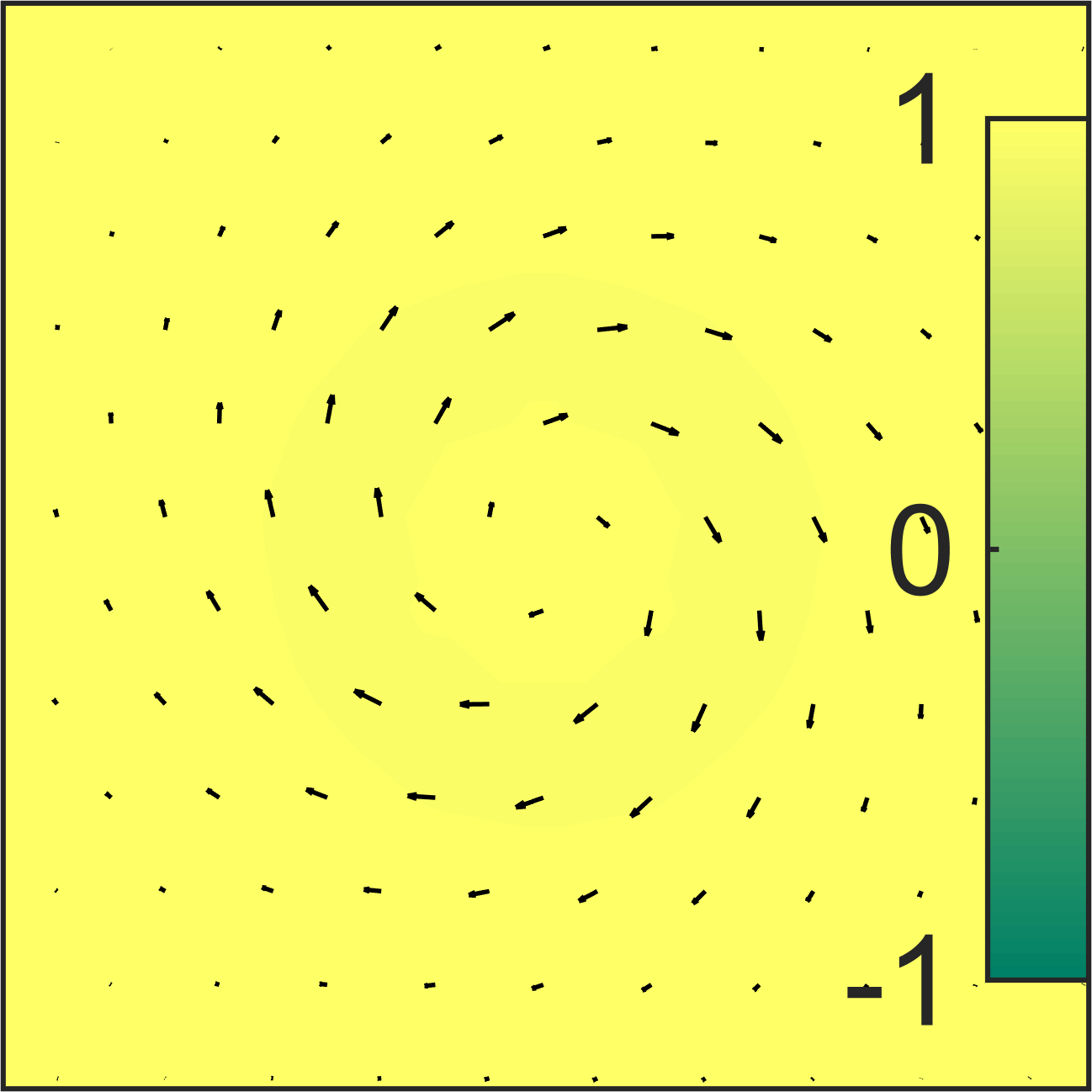}\\
	\begin{centering}
		\caption{\label{fig:FMAFMskyrmion}
Spin configurations in the layers containing the skyrmion in the (a) A-type AFM state and  (b) FM state. (c) and (d) The corresponding spin configurations in 
neighboring layers.
In-plane spin components are shown with arrows; the out-of-plane spin components are color coded.}
		\par\end{centering}
\end{figure}

Although the AFM interlayer coupling suppresses skyrmion crystals, it allows for isolated skyrmions with topological charge confined to a single layer.  
Remarkably, these single-layer skyrmions are stable in three different phases: the FM, AFM and CS states in the region  encircled by red dotted line in Fig.~\ref{fig:PHD}.
Figure~\ref{fig:FMAFMskyrmion}a shows the skyrmion in the AFM state, which is only stable in layers with the magnetization opposite to the applied field. 
The magnetic moment in the skyrmion center is  parallel to the field, in contrast to FM skyrmions (Fig.~\ref{fig:FMAFMskyrmion}b). 
The small disturbance  in the spin configuration of neighboring layers is topologically trivial (Fig.~\ref{fig:FMAFMskyrmion}c) and its magnitude decays quickly with the distance from the skyrmion (see Supplementary information).
It has a stabilizing effect on skyrmions which are unstable in the surface layers. 

\begin{figure}
	\raggedright{\bf\large a \hskip 4.cm b}\\
	\includegraphics[height=0.49\columnwidth]{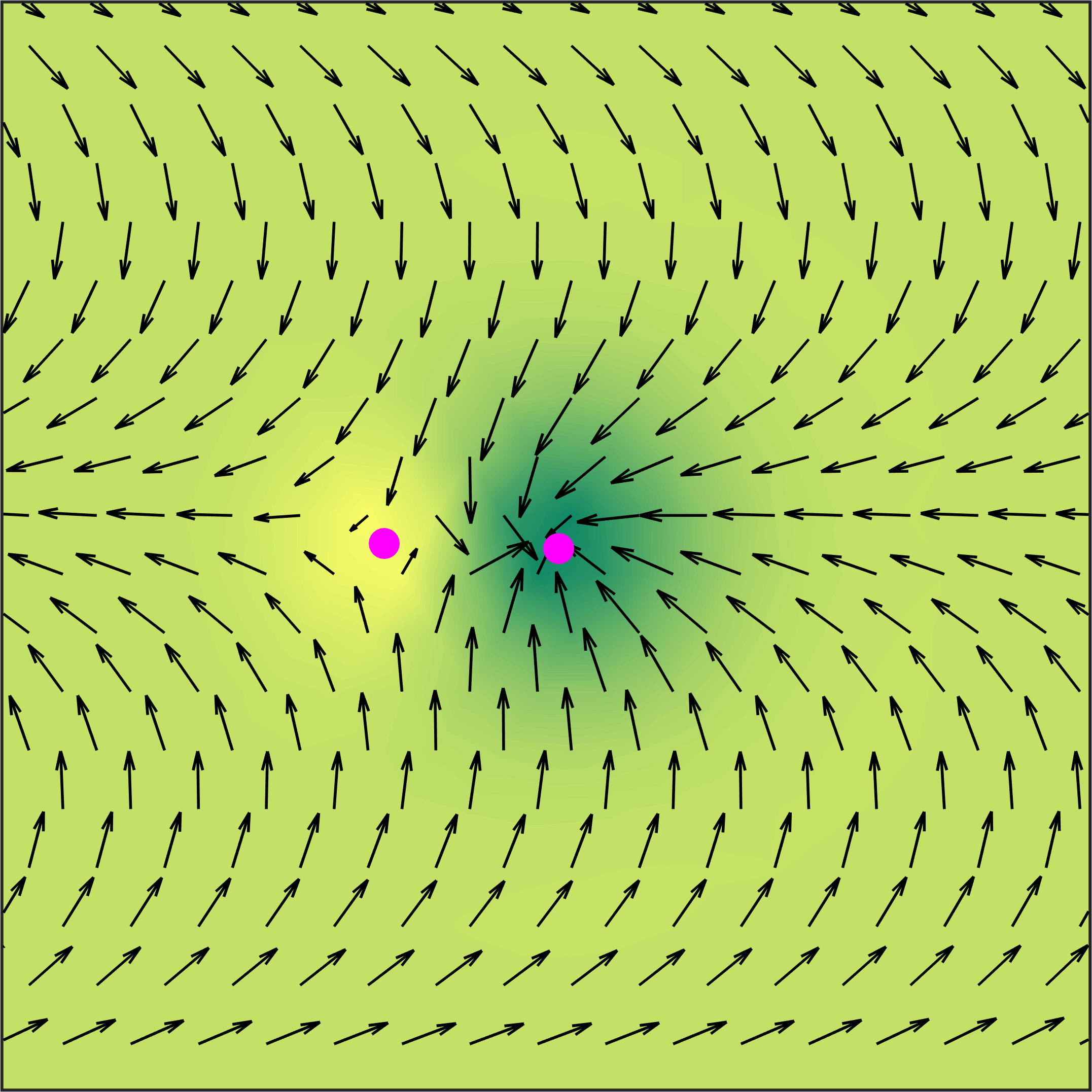}\hfill
	\includegraphics[height=0.49\columnwidth]{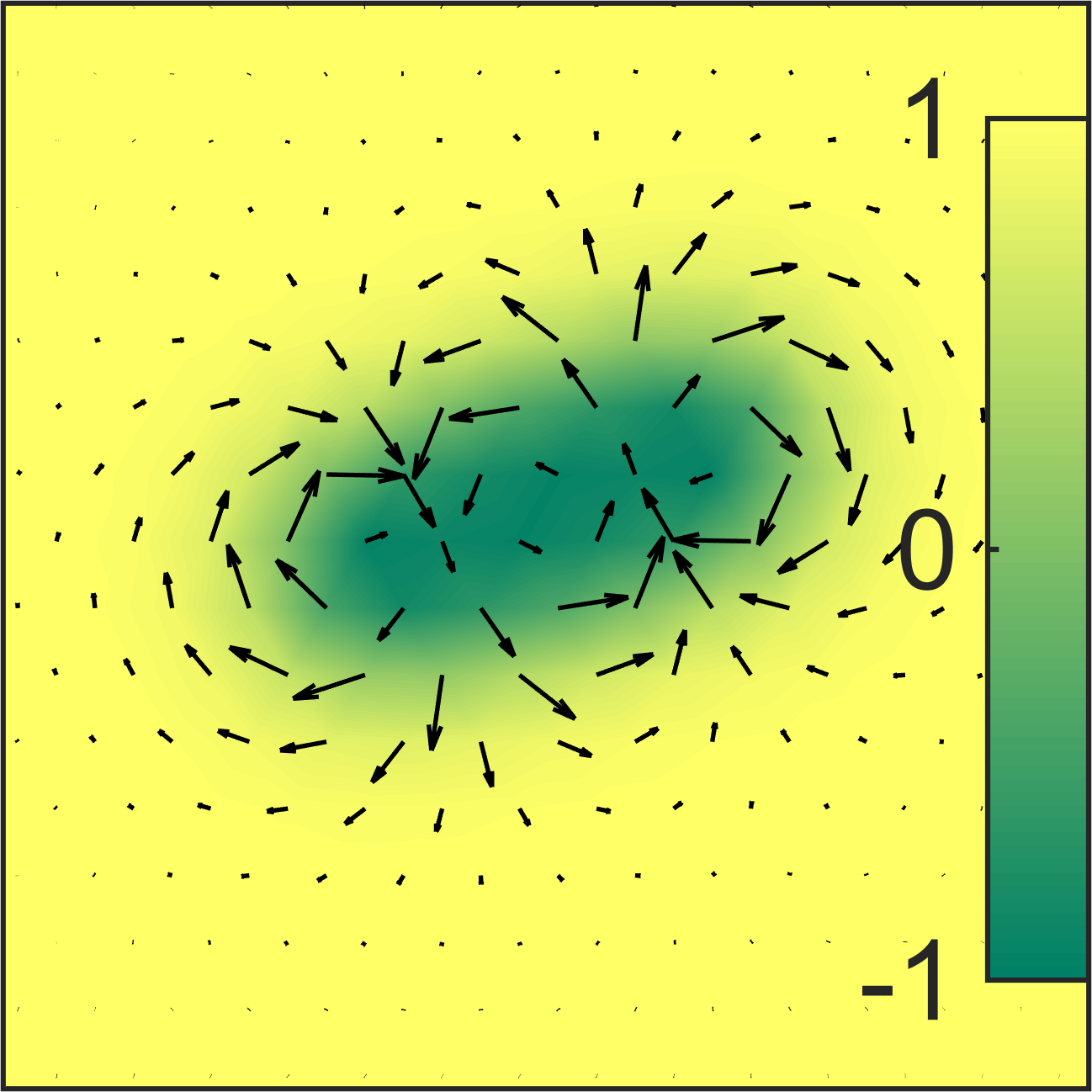}\\
	\raggedright{\bf\large c \hskip 4.cm d}\\
	\includegraphics[height=0.49\columnwidth]{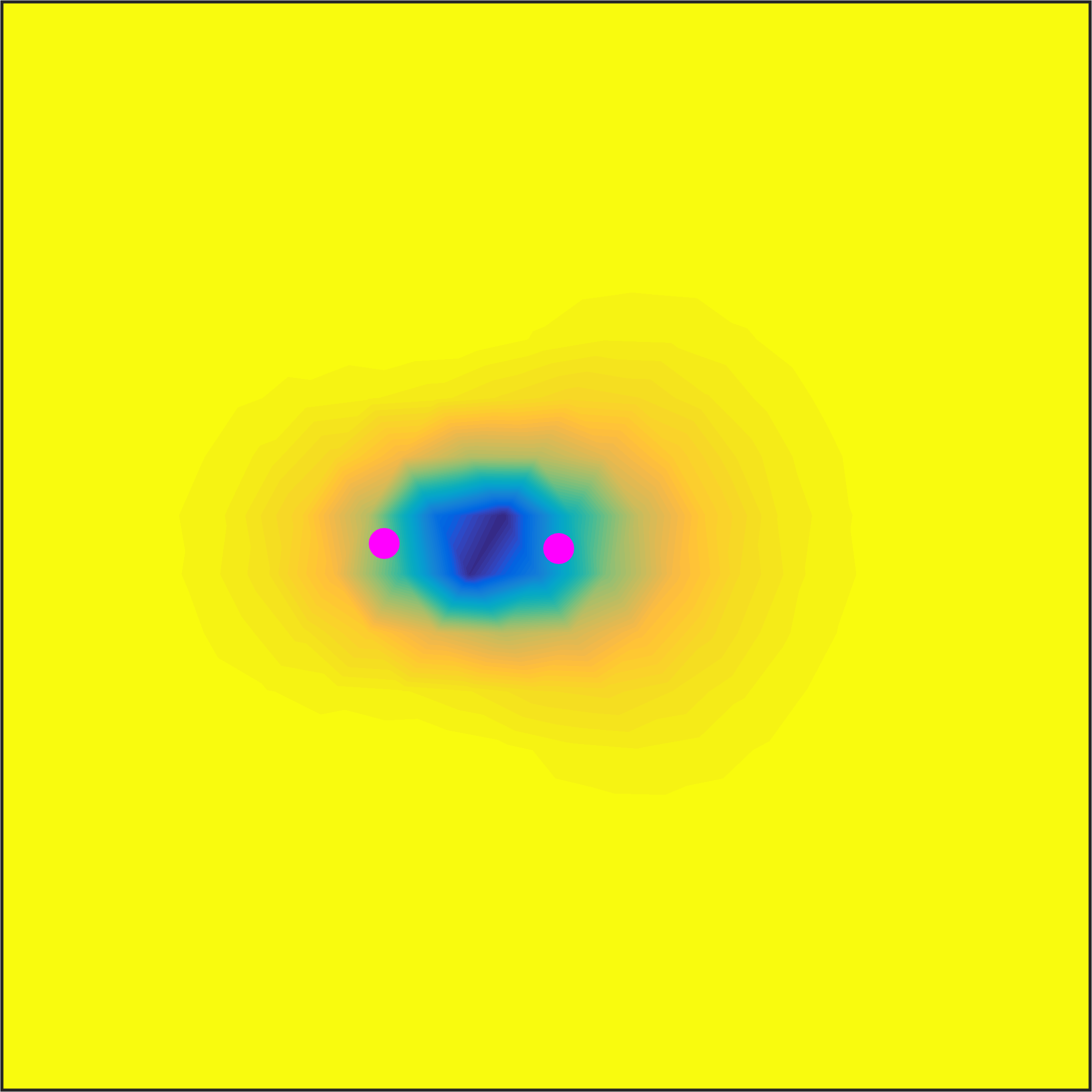}\hfill
	\includegraphics[height=0.49\columnwidth]{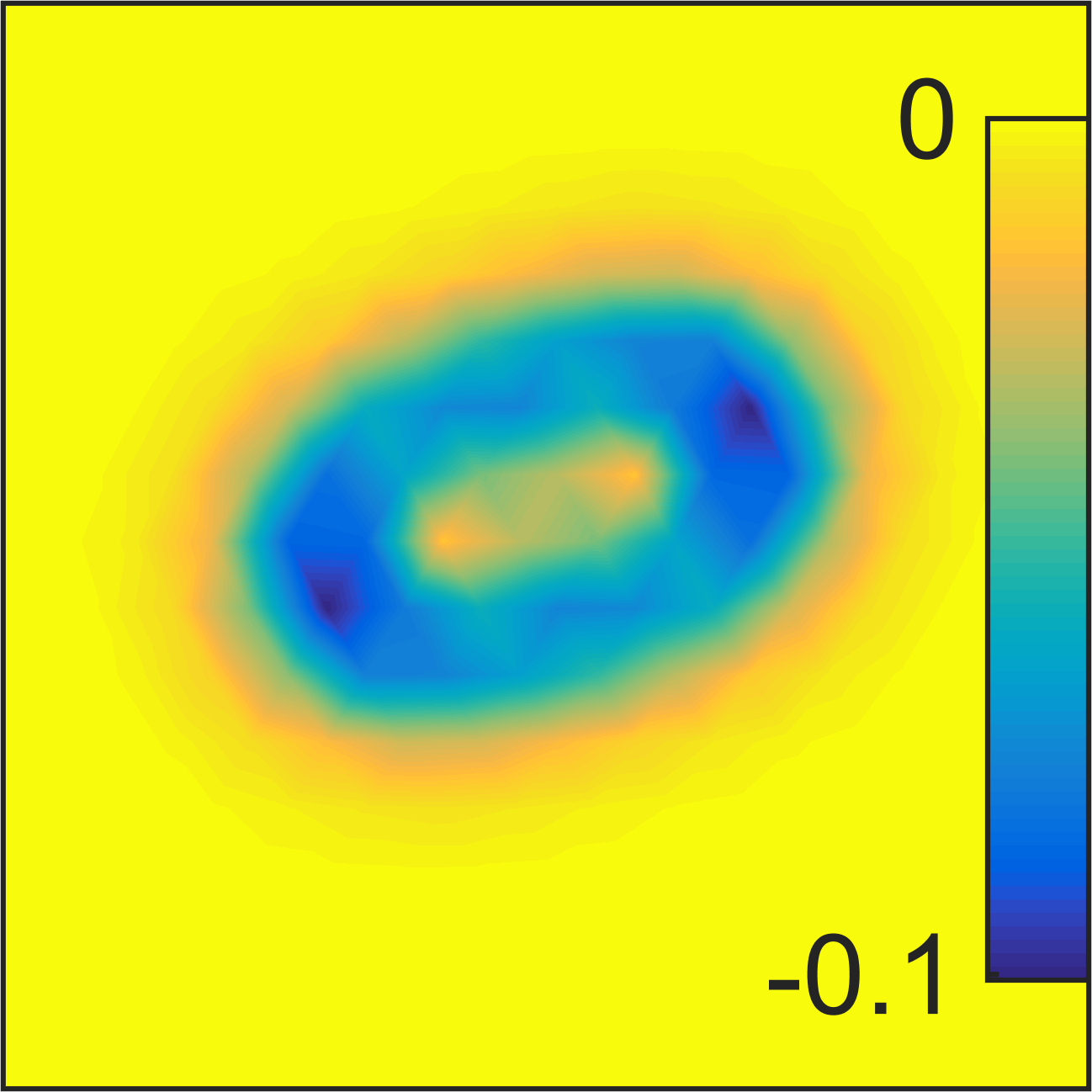}\\
	\begin{centering}
		\caption{
			\label{fig:skyrmion_CS_Q23}		
(a) Bi-meron in the CS state, (b) skyrmion with topological charge 3 in the FM state, and the corresponding distributions of the topological charge density  (panels c and d).
In-plane spin components are shown with arrows; the out-of-plane spin components and the topological charge density are color coded. Pink dot marks the (anti)vortex center.}
		
		\par\end{centering}

\end{figure}

Unexpectedly,  a magnetic defect with skyrmion topology is also stable in the CS state (Fig.~\ref{fig:skyrmion_CS_Q23}a). 
It has the appearance of a bi-meron -- the bound vortex-antivortex pair~\cite{Kharkov2017}, but its topological charge distribution has a single maximum and is not divided into two halfs  (see Fig.~\ref{fig:skyrmion_CS_Q23}c).
We  also found single-layer skyrmions with topological charges $Q= \pm 2$ and $\pm 3$ in the FM state (Figs.~\ref{fig:skyrmion_CS_Q23}b,c) stabilized by AFM interlayer interactions.

{\flushleft\it Dynamics:}
Next we discuss dynamics of single-layer skyrmions induced by the spin-Hall torque \cite{Slonczewski1996,Khvalkovskiy2013}
\begin{equation}
\label{eq:spin-Hall_torque}
\mathbf{T}_{\rm SH} \propto \mathbf{S}\times \mathbf{S} \times 
\left(\hat{\mathbf{c}} \times \mathbf{j}_{\rm c} \right),
\end{equation} 
where {\bf j$_{\rm c}$} is the charge current in a heavy metal layer  generating a spin accumulation at the interface with the antiferromagnet, $\mathbf{S}$ is a spin in the interface layer (layer 1) and the unit vector $\hat{\mathbf{c}}$ is normal to spin layers.
Due to the instability of skyrmions at the interface, we study their dynamics in layer 2 (next to the interface layer) by numerical integration of Landau-Lifshitz-Gilbert equation with the torque (see Suplementary information).

\begin{figure}
	\raggedright{\bf \large a}\\
	\centering{}\includegraphics[width=\columnwidth]{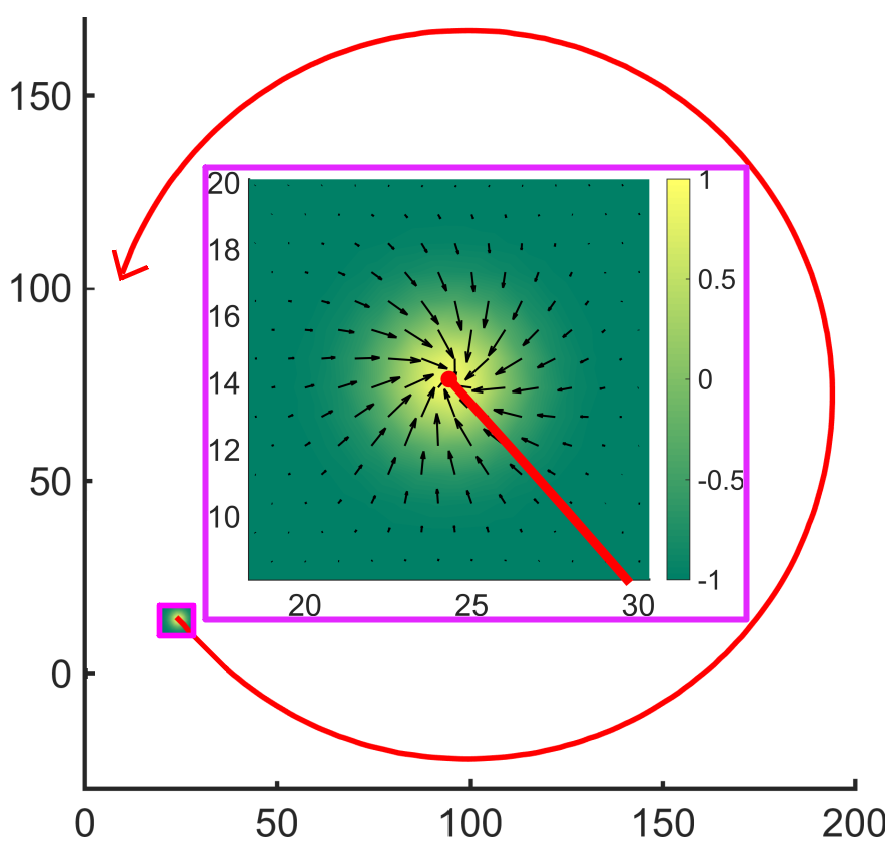}\\
	\raggedright{\bf\large  b}\\
	\centering{}\includegraphics[width=\columnwidth]{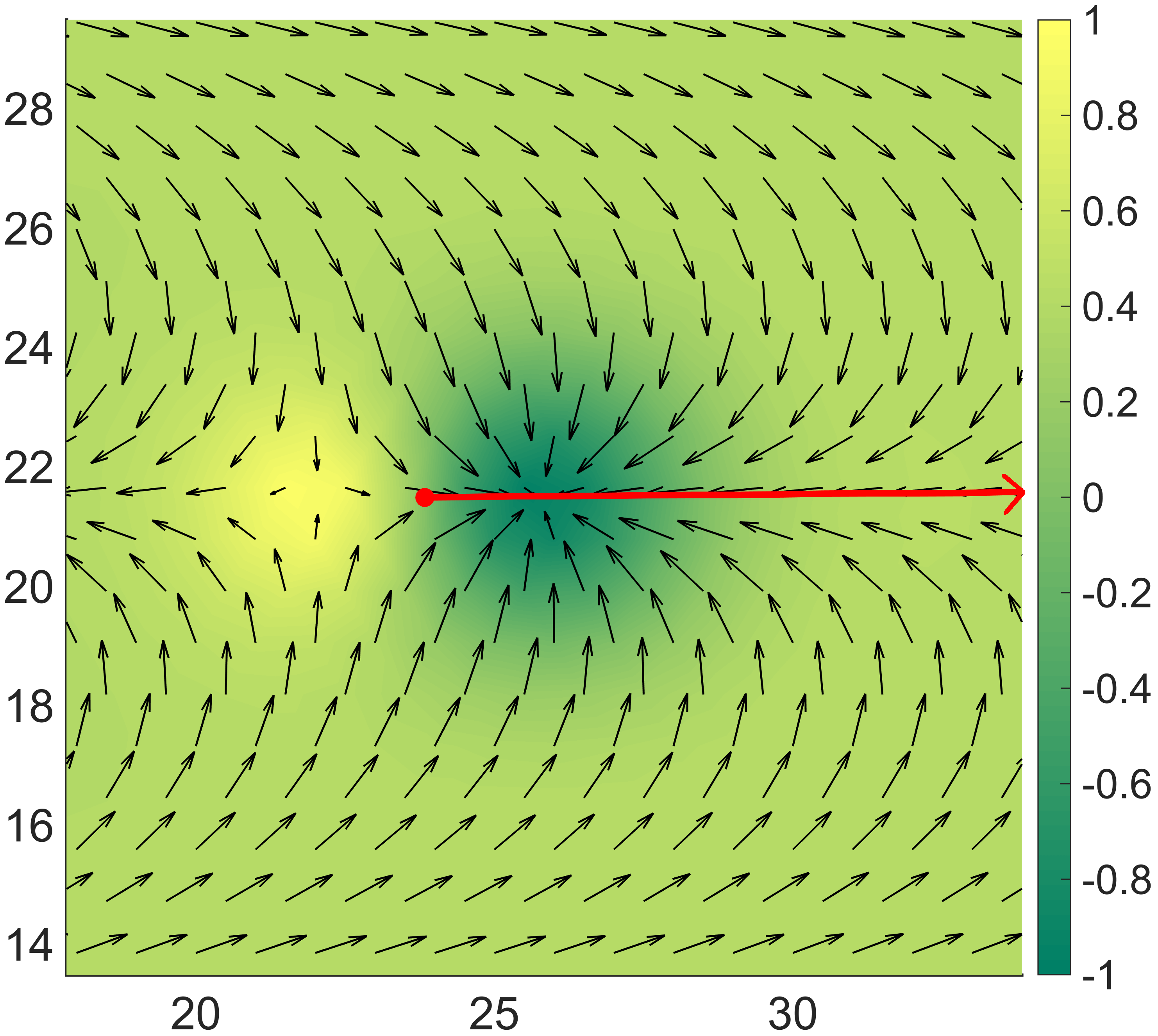}\\
	\begin{centering}
		\caption{
			\label{fig:trajectories}
Trajectories of topological defects in  (a) the A-type 
AFM and (b) CS states moved by the spin-Hall current applied along the positive $x$ direction.}
		
		\par\end{centering}

\end{figure}

The skyrmion in the uniform AFM and FM states shows the coupled helicity and center-of-mass dynamics \cite{Leonov2017} with  the helicity angle growing linearly with time, corresponding to  rotation of the in-plane spins, and the rotational center-of-mass motion \cite{Lin2016,Zhang2017} (see Fig.~\ref{fig:trajectories}a and  Supplementary Movie 1). 
Surprisingly, the bi-meron motion in the CS state only depends on the direction of ${\bf j}_{\rm c}$ in the initial stage. 
In the steady state, the bi-meron moves along a straight line perpendicular to the spiral wave vector $\mathbf{Q}$ and its helicity remains constant (see Fig.~\ref{fig:trajectories}b  and Supplementary Movies 2--3).
The spin-Hall torque results in a sinusoidal modulation of $S^z$ in layer 1 and the interlayer interaction attracts bi-merons into a `channel' perpendicular to ${\bf Q}$, located under the $S^z \approx + 1$ line in layer 1.
The channel position and the skyrmion helicity do depend on ${\bf j}_{\rm c}$, but the direction of motion does not.
This peculiar dynamics is a consequence of the rotational invariance of Heisenberg exchange interactions and the fact that the rotation of in-plane spins is equivalent to a translation of the CS along ${\bf Q}$. 
The transverse motion of bi-merons is accompanied by a slow drift along  ${\bf Q}$, which is a finite-size artifact that becomes smaller as the system size increases (see Supplementary information).  

%

%As a result of gapped modes due to Skyrmion inner deformations during excitation
% \cite{Lin2017},
%its helicity varies approximately linearly with time (ref. to figure). 
%Helicity and skyrmion velocity are coupled, so that when the skyrmion starts moving 
%in a certain direction, its helicity varies as a consequence. In turn the different helicity 
%results in a different velocity direction. 

%

%
%The sign of the skyrmion velocity depends on the sign  of $\mathbf{J}_{\rm SH} \cdot [ \hat{\mathbf{c}} \times \mathbf{Q}]$.
%
%For $J_{\rm SH} || x$ and  $Q || y$, the skyrmion moves along the positive $x$ direction.
%as shown in Fig. \ref{fig:dynamics}.

%Here, where $\mathbf{J}_H = J_H\hat{x}$,
%it can be seen that there is a small $y$ component to the defect velocity which is due to the translation of the shole spin configuration. 
%
%This effect should not be confused for the Magnus force typically present  in skyrmion dynamics and turns out to be dependent on the size of our simulation: it would therefore not occur in a macroscopic system since its magnitude
%decreases with the increasing size (see figure).

{\flushleft\it Conclusions:}\,
In conclusion, centrosymmetric NiBr$_2$ has a number of properties required to host skyrmion lattices: frustrated exchange interactions, three-fold symmetry axis and magnetic anisotropy tunable by Fe-doping. 
However,  because of the long magnetic modulation period, even the relatively weak AFM interlayer interactions suppress the skyrmion crystal and other multiply-periodic states in this material.
%
%In this respect, more frustrated di-halides showing spiral states with a shorter period, such as NiI$_2$ \cite{Kuindersma1981}, have a better chance to host skyrmion crystals.
%
Yet the interlayer interactions do not prevent the formation of isolated single-layer skyrmions and even have a stabilizing effect on them.
 The single-layer skyrmions are mobile, stable in several magnetic phases and, in contrast to AFM skyrmion tubes, can be easily nucleated, which makes them promising candidates for information carriers in antiferromagnetic spintronics.

\end{document}